\documentclass{article}  
\usepackage{spadre2008}
\usepackage{graphicx}
\usepackage{amssymb}
\usepackage{amsmath}
\usepackage{amsthm}
\usepackage{epsf}
\frompage{000} \topage{000}                                              %|
\title{Understanding Jet Energy Loss with \\ 
Angular Correlation Studies in PHENIX} 
\authors{
{J. A. Hanks for the PHENIX Collaboration$^1$ }\\[2.812mm]
{\normalsize
\hspace*{-8pt}$^1$ Department of Physics, Columbia University, \\ 
New York, NY 10027, USA\\[0.2ex] 
}}
 
\abstract{Angular correlation studies provide powerful insight into the 
energy loss of hard scattered partons as they traverse the partonic medium 
produced in heavy ion collisions at RHIC. These results are generally compared 
to jet correlations in p+p collsisions where all correlation strength is 
attributed to vacuum fragmentation. Strong modification to di-jet correlations 
has been observed in A+A collisions at RHIC, most notably for the away side 
jet. Many different effects, including the opacity of the medium, its response 
to energy deposited by partons as they propagate, and modifications to the 
parton fragmentation, are involved in producing the final correlation 
stuctures. Understanding the interplay between these various effects is 
essential to developing a complete picture of the medium. Measurements of jet 
correlations involving direct photons provide a unique probe of jet 
fragmentation effects, as photons are not strongly interacting. Additionally, 
systematic studies of the away side structure as a function of $p_{T}$, as 
well as attempts to include additional high $p_{T}$ trigger requirements, can 
help to distinguish different energy loss mechanisms. We discuss recent 
PHENIX results from these detailed studies of jet correlations in A+A and p+p 
collisions.
}

\keyword{heavy ions, correlations, jets} 
\PACS{25.75.Gz}
 
\begin{document}
 
\maketitle
\setcounter{page}{1}

\section{Introduction}\label{intro}

In heavy ion collisions at RHIC we observe a significant modification of jets 
as a result of their interaction with the medium \cite{bib1}. We are able to measure 
such modifications through two particle angular correlations, in which the $\Delta\phi$
between a trigger particle and all associated particles is measured and the large 
background can be dealt with using statistical subtraction methods \cite{bib2}. These 
correlation measurements are performed over a wide range of trigger and associated 
particle momenta, allowing for a detailed study of how medium modifications depend on both.

The modification of the away-side jet appears in two ways. The first is in the form of a 
supression of the di-jet at $\Delta\phi = \pi$, which we will refer to as the the "head" region. This 
suppression can be described well at high $p_{T}$ in terms of energy loss \cite{bib3}. The 
second is in the form of an enhancement away from $\pi$, in the "shoulder" region, thought 
to be due to medium response. There are several models which attempt to explain how this 
enhancement is produced. One possibility is that the jet propogates faster than the speed 
of sound in the medium, producing a Mach cone \cite{bib4}. Another is that the jet is 
propagating faster than the speed of light in the medium, producing Cherenkov radiation 
\cite{bib5}. Finally, the jet propagation may be coupled with the collective flow, causing 
it to be deflected away from $\pi$ \cite{bib6}. These various scenarios can be tested by 
studying the position of the shoulder peaks as a function of $p_{T}$ and centrality \cite{bib3}, 
as well the reaction plane dependence.

Much information can be gained from these di-hadron studies, however there are inherent 
limitations to how well energy loss can be understood through these measurements, both as a 
result of the surface bias introduced by the trigger requirement \cite{bib7}, as well as the 
fact that both the trigger and associated hadrons will experience modification due to the 
medium. These difficulties can, at least in part, be overcome by measuring correlations of 
hadrons with direct photons, which do not interact with the medium.

\section{Away-side Jet Modification}\label{awayside}

There are several methods in PHENIX for studying the modification of the away-side jet. To 
understand the interplay between the head and shoulder regions the away-side can be decomposed 
into three sections: the head region, centered around $\Delta\phi = \pi$ and ranging from 
$\pi - \pi/6$ to $\pi + \pi/6$; and the two shoulder regions, in the range 
$\pi/6 < |\Delta\phi - \pi | < \pi/2$. This allows us to study how these various regions evolve. 
To measure the positions of the shoulder regions there are two fitting procedures that are used. 
The first uses two gaussians diplaced from $\pi$ by a distance D, as shown in equation (1).

\begin{equation}
J(\Delta\phi)_{away-side} = A(e^{\frac{(\Delta\phi -\pi + D)^{2}}{2\sigma^{2}}} + 
e^{\frac{(\Delta\phi -\pi - D)^{2}}{2\sigma^{2}}})
\end{equation}

The second includes a third gaussian, representing the head contribution, which is fixed by the 
p+p width, and centered at $\pi$, as shown in equation (2).

\begin{equation}
J(\Delta\phi)_{away-side} = A(e^{\frac{(\Delta\phi - \pi + D)^{2}}{2\sigma^{2}}} + 
e^{\frac{(\Delta\phi - \pi - D)^{2}}{2\sigma^{2}}})
+ Be^{\frac{(\Delta\phi - \pi)^{2}}{2\sigma_{pp}^{2}}}
\end{equation}

\begin{figure}[htb]
\begin{center}
$\begin{array}{cc}
 \includegraphics[width=5.0cm]{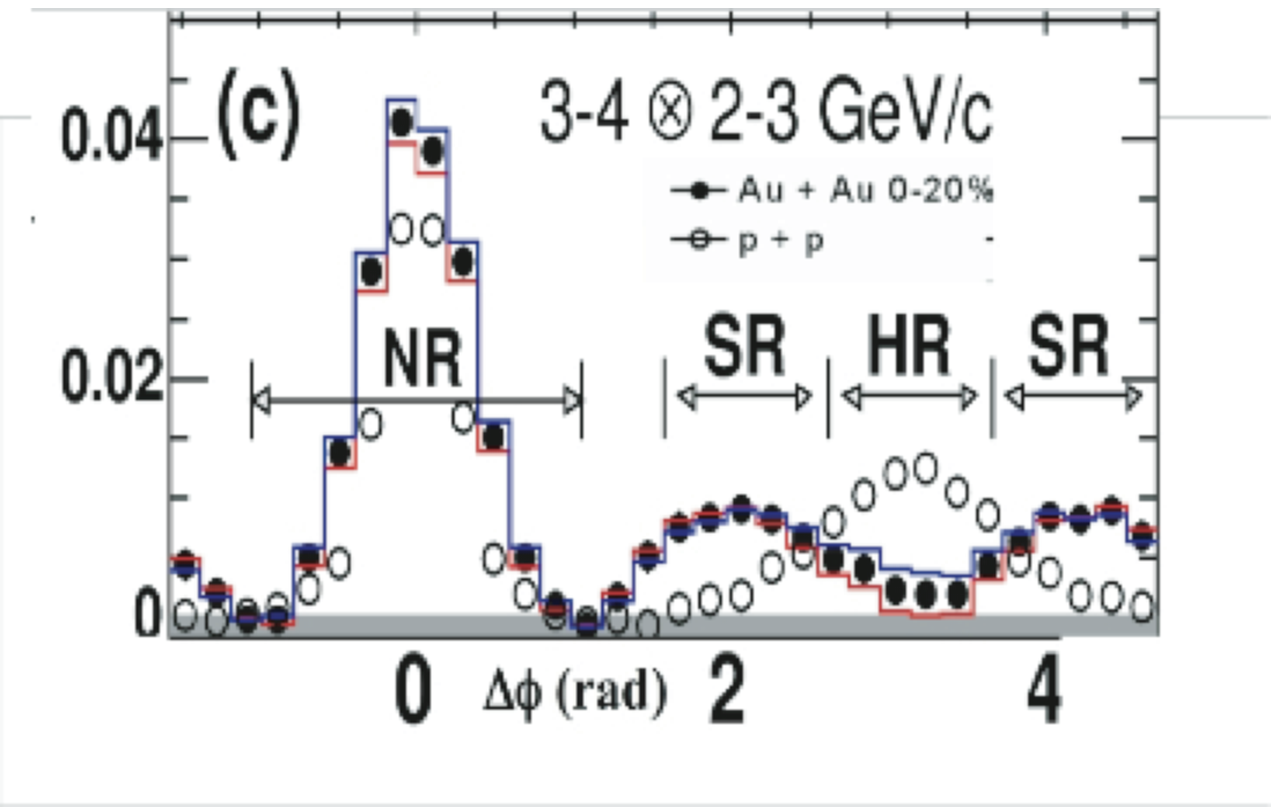} &
 \includegraphics[width=6.5cm]{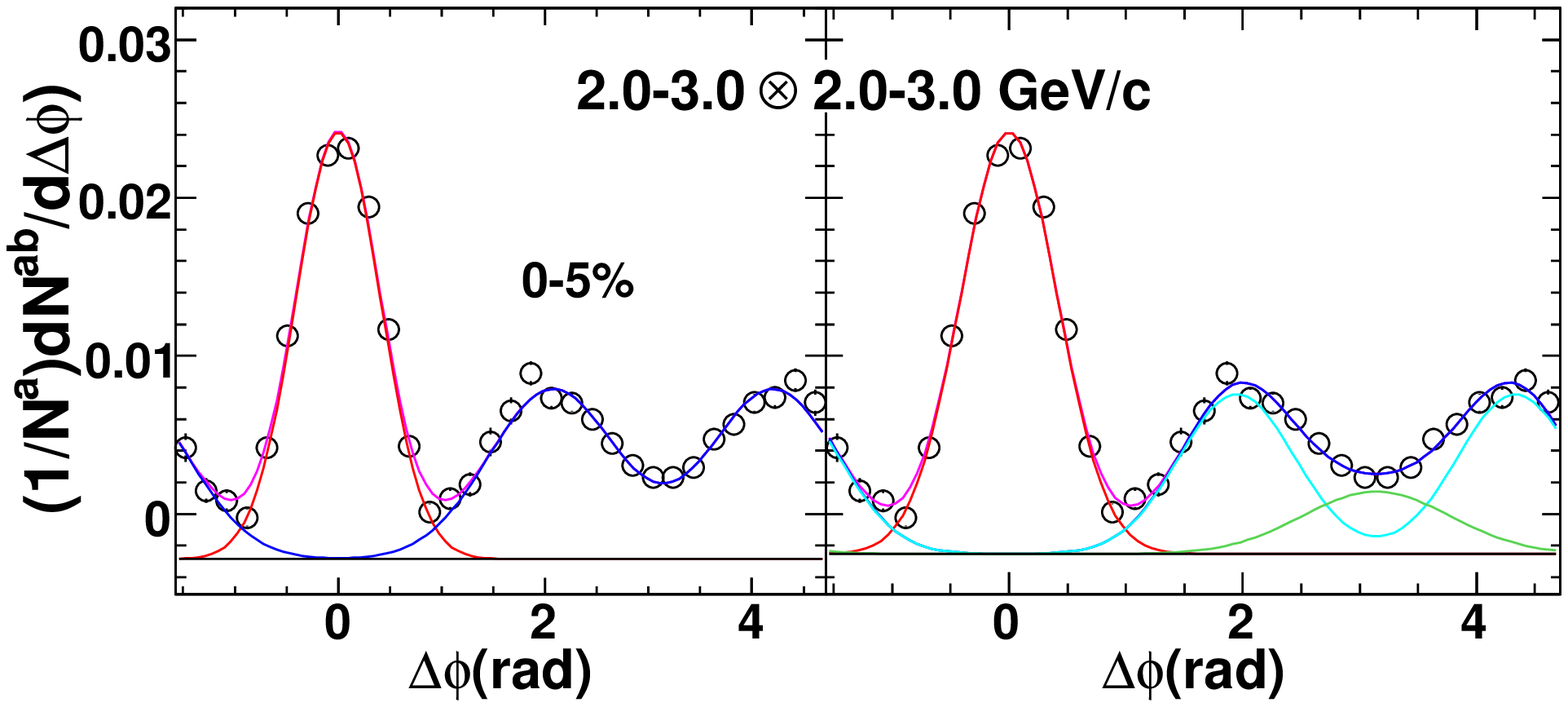} \\
\mbox{\bf (a)} & \mbox{\bf (b)}
\end{array}$
\caption{{\bf(a) } Decomposition of away-side into two seperate components, the head and the shoulder.
 {\bf(b)} Demonstration of the two fits used to determine seperation of two shoulder peaks.}
\label{fig:decomp}
\end{center}
\end{figure}

In these equations J is used to indicate that it is the jet yield being measured. An illustration of both the head/shoulder decomposition and the two fitting methods is shown in 
figure \ref{fig:decomp}.

Studying the dependence of the head and shoulder regions on both the trigger and associated 
$p_{T}$ reveals that there is an enhancement of the shoulder yield at low associated $p_{T}$ 
which persists as we go to higher trigger momentum. This low $p_{T}$ enhancement gives the 
shoulder a softer spectral shape, making it appear more bulk-like. We also find that the 
head is suppressed at intermediate to high associated $p_{T}$, for all trigger $p_{T}$ bins 
\cite{bib3}. Using the decomposition of the head and shoulder regions also allows us to compare 
them directly by looking at the ratio of the integrated yields, $R_{HS}$, as described by equation (3).

\begin{equation}
R_{HS} = (\int_{head}d\Delta\phi \frac{1}{N_{trig}}\frac{dN}{d\Delta\phi})/
(\int_{shoulder}d\Delta\phi \frac{1}{N_{trig}}\frac{dN}{d\Delta\phi})
\end{equation}

We find that the shoulder dominates at low partner momentum, up to $p_{T} \approx 4$~GeV/c, 
while as the trigger $p_{T}$ is increased this ratio becomes consistent with the p+p measurement 
when the partner $p_{T}$ is greater than $4$~GeV/c \cite{bib3}.

\begin{figure}[htb]
\begin{center}
 \includegraphics[width=5.5cm,height=6.0cm]{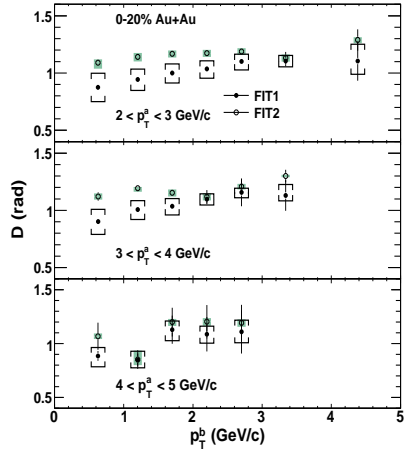}
\caption[]{ Shows separation between two shoulder peaks, in the form of the D fit parameter. The fit
 described by equation (1) is shown in black, the one described by equation (2) is in green.}
\label{fig:shoulder_position}
\end{center}
\end{figure}

The two fitting methods can be used to extract the D parameter, which is a measure of the 
separation between the shoulder peaks, and study its dependence on the momentum of either 
hadron. We find that the peak positions show minimal dependence on either the trigger or 
partner $p_{T}$, as shown in figure \ref{fig:shoulder_position} \cite{bib3}. This poses a 
challenge for models such as Cherenkov radiation which predict a $p_{T}$ dependence.

It is also possible to use particle ID to measure these types of correlations for baryons and 
mesons separately to determine how the modification observed depends on species. The results show 
a similar shape to the away side correlations for both. Additionally, the ratio of baryons to 
mesons in the away side yields shows a similar centrality dependence to that of the bulk, which 
is incompatible with vacuum fragmentation \cite{bib8}. These observations place a tight constraint 
on models attempting to describe baryons and mesons together, and provides further evidence for the 
idea that the shoulder region is dominated by medium response.

\section{$\gamma$-h and h-$\gamma$}\label{photons}

Correlations between hadrons and photons can go even futher to improve our understanding of 
the various affects leading to both the energy loss of the jet, as well as the medium response, 
since the photons remain unmodified. In $\gamma$-jet correlations the near-side parton essentially 
fragments all of its energy into the photon via Compton scattering; the photon then emerges without 
further interaction. This allows us to measure the initial energy of the recoil jet directly by measuring 
the photon energy, and therefore to measure the amount of energy loss for the parton in the medium. 
Additionally, di-jet measurements will be dominated by jets produced near the surface where the 
energy loss effects are minimal, leading to a geometric bias. The use of a photon trigger removes 
this bias, allowing for a measurement of the full geometry of the medium through which the jet is 
propogating.

\begin{figure}[htb]
\begin{center}
$\begin{array}{cc}
 \includegraphics[width=5.5cm]{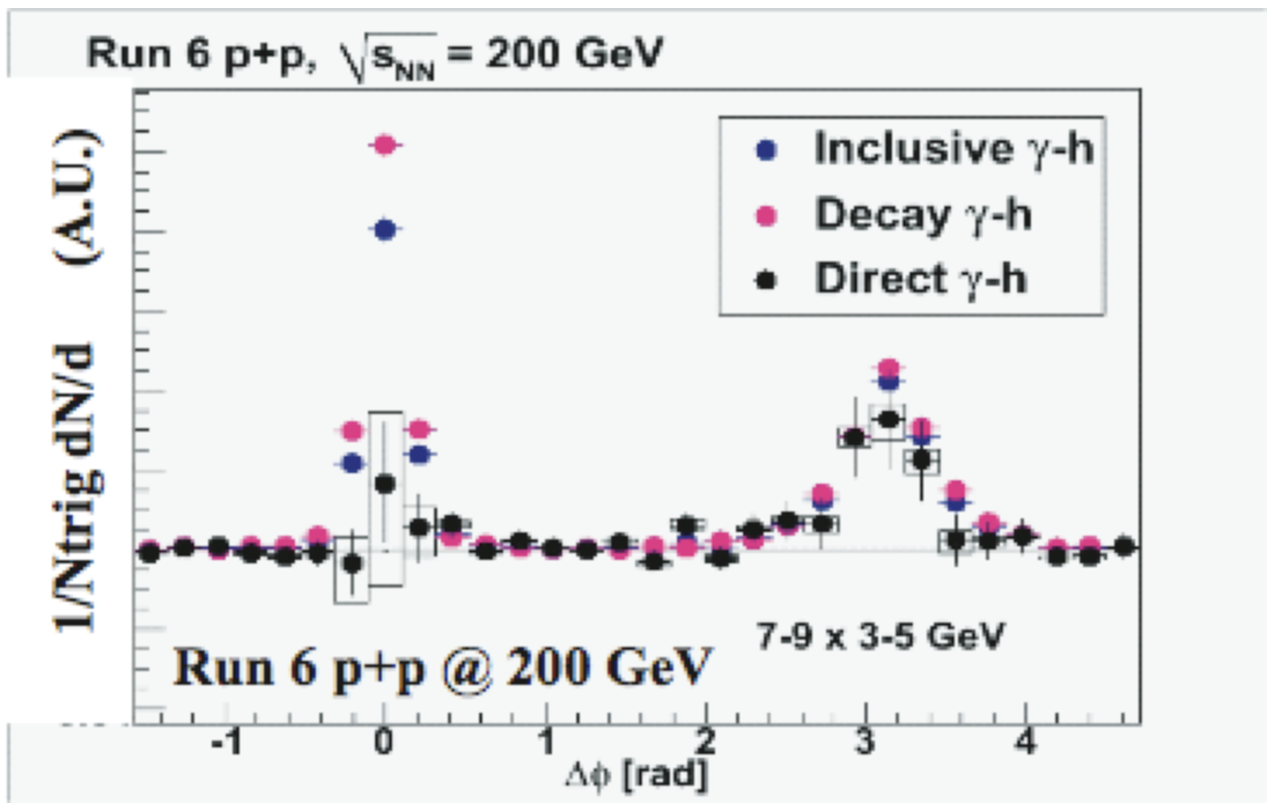} &
 \includegraphics[width=3.5cm]{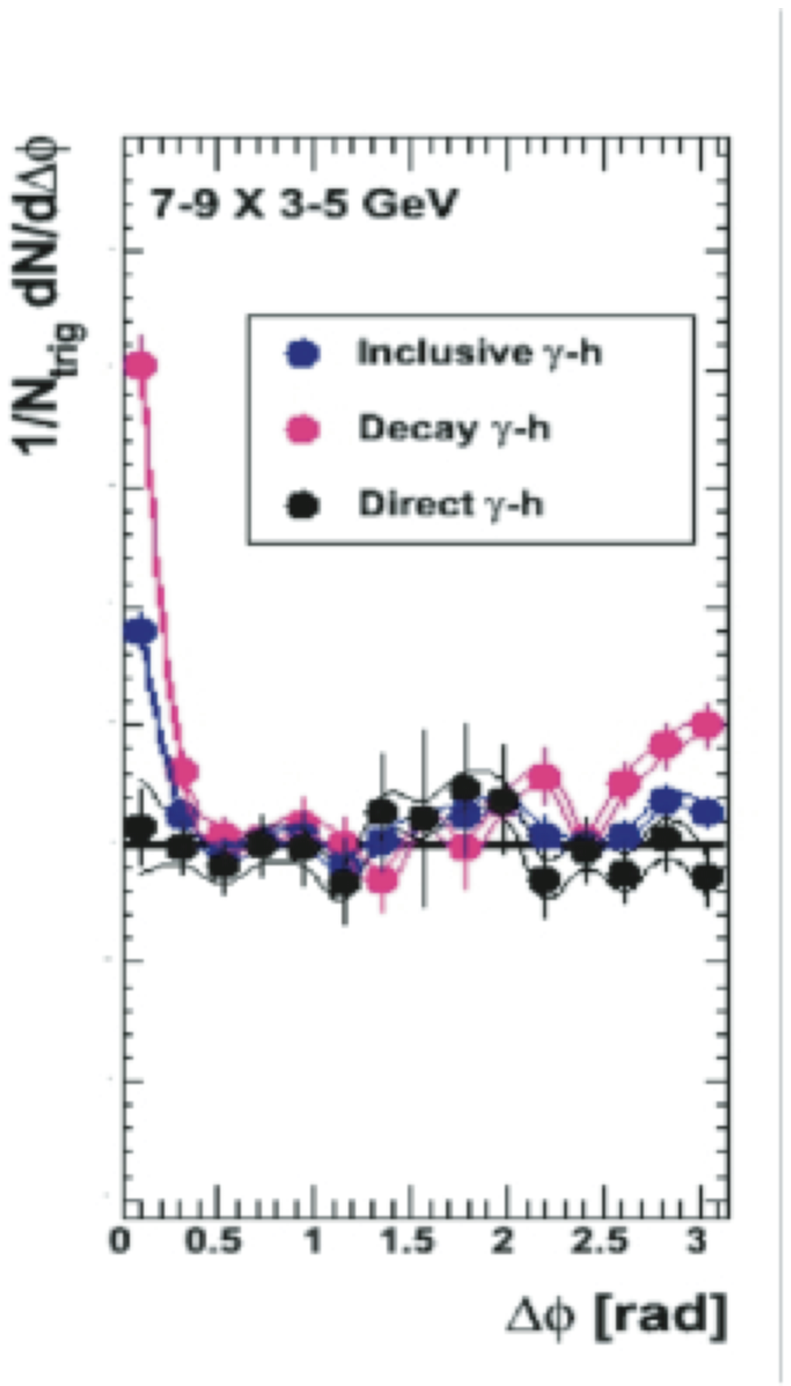} \\
\mbox{\bf (a)} & \mbox{\bf (b)}
\end{array}$
\caption[]{{\bf(a) } $\gamma$-hadron per trigger yields in p+p collisions: inclusive(blue), decay(red), and direct(black). {\bf(b)} $\gamma$-hadron per trigger yields in Au+Au collisions: inclusive(blue), decay(red), and direct(black).}
\label{fig:gamma_jet}
\end{center}
\end{figure}

Figure \ref{fig:gamma_jet} shows the measured correlation functions both in p+p and Au+Au. In the 
case of p+p, the near-side yield is consistent with zero, as is expected if the photon carries 
the full parton energy, and the away-side yield is similar to that for the case of $\pi^{0}-h$ 
correlations. In the case of Au+Au there is a clear suppression relative to the p+p, indicating that 
this measurement is sensitive to the energy loss of the away-side jet. These preliminary results 
are encouraging for future more detailed studies of the modification to the away-side via photon-jet 
correlations.

\begin{figure}[htb]
\begin{center}
 \includegraphics[width=8.5cm,height=5.0cm]{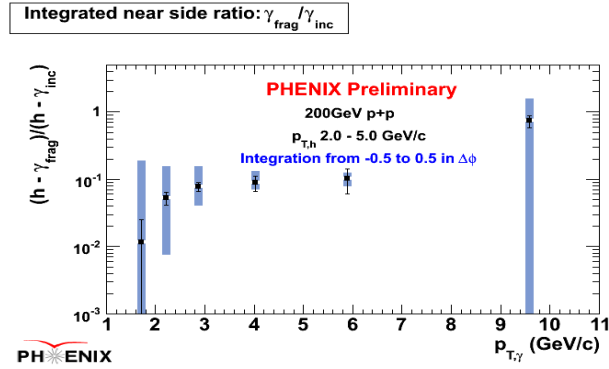}
\caption[]{ Shows the ratio of integrated near-side per trigger yield, from $-0.5$ to $0.5$ in $\Delta\phi$
for fragmentation photons relative to inclusive.}
\label{fig:frag_photons}
\end{center}
\end{figure}

While the near-side yield in the $\gamma$-jet measurement is consistent with zero, there is expected 
to be some non-zero yield as a result of direct photons produced as the parton fragments. In addition, 
the production of fragmentation photons in heavy ion collisions is expected to be modified as a result 
of the interaction of the jet as it propagates. Direct measurement of this component of the direct 
photon signal can be attempted through h-$\gamma$ correlation measurements, as these photons should be 
correlated with the jet which produces them. A first attempt to perform this measurement in p+p 
collisions has been made, and figure \ref{fig:frag_photons} shows the ratio of the near-side integrated 
yields for fragmentation photons over inclusive. This measurement is still dominated by large 
systematics, but a yield of 5-15$\%$ is observed at intermediate $p_{T}$, and the success of the method 
is encouraging for future improvements, as well as possible measurements in larger systems.

\section{Conclusions}\label{concl}

Jet correlations provide a powerful tool for probing energy loss in heavy ion collisions. Studies of two 
particle correlations over a wide range of $p_{T}$, over all centralities, and as a function of reaction 
plane have all been done at PHENIX. The results provide a description of jet energy-loss and medium response 
consistent with the idea that medium response dominates at intermediate $p_{T}$, and the dependence of this 
medium response on various parameters such as the trigger and partner $p_{T}$'s, and reaction plane angle 
are not inconsistent with models describing jet induced Mach cones, but pose a challenge for other scenarios 
such as Cherenkov gluon radiation.

\vfill\eject

\end{document}